\documentclass[twocolumn,amsmath,amssymb,superscriptaddress]{revtex4}

\usepackage{graphicx}
\usepackage{dcolumn}
\usepackage{bm}

\begin{document}


\title{Hoffmann-Infeld Black Hole Solutions in Lovelock Gravity}

\author{Mat\'\i as Aiello}
\email{aiello@iafe.uba.ar}
\affiliation{Departamento de F\'\i sica, Facultad de Ciencias Exactas y
Naturales, Universidad de Buenos Aires, Ciudad Universitaria, Pabell\'on
I, 1428 Buenos Aires, Argentina}
\affiliation{Instituto de  Astronom\'\i a y F\'\i sica del Espacio,
C.C. 67, Suc. 28, 1428 Buenos Aires, Argentina}
\author{Rafael Ferraro}
\email{ferraro@iafe.uba.ar}
\thanks{Member of Carrera del Investigador Cient\'{\i}fico (CONICET,
Argentina)}
\affiliation{Departamento de F\'\i sica, Facultad de Ciencias Exactas y
Naturales, Universidad de Buenos Aires, Ciudad Universitaria, Pabell\'on
I, 1428 Buenos Aires, Argentina}
\affiliation{Instituto de  Astronom\'\i a y F\'\i sica del Espacio,
C.C. 67, Suc. 28, 1428 Buenos Aires, Argentina}
\author{Gast\'on Giribet}
\email{gaston@df.uba.ar}
\thanks{Member of Carrera del Investigador Cient\'{\i}fico (CONICET,
Argentina)} \affiliation{Departamento de F\'\i sica, Facultad de
Ciencias Exactas y Naturales, Universidad de Buenos Aires, Ciudad
Universitaria, Pabell\'on I, 1428 Buenos Aires, Argentina}
\affiliation{Departamento de F\'{\i}sica, Universidad Nacional de
La Plata, C.C. 67, 1900 La Plata, Argentina.}
\begin{abstract}

Five-dimensional black holes are studied in Lovelock gravity
coupled to Hoffmann-Infeld non-linear electrodynamics. It is shown
that some of these solutions present a double peak behavior of the
temperature as a function of the horizon radius. This feature
suggests that the evaporation process, though drastic for a period,
leads to an eternal black hole remnant. In fact, the form of the
caloric curve corresponds to the existence of a {\it plateau} in the evaporation rate,
which implies that black holes of intermediate scales turn out to
be unstable. The geometrical aspects, such as the absence of
conical singularity, the structure of horizons, etc. are also
discussed. In particular, solutions that are asymptotically AdS
arise for special choices of the parameters, corresponding to
charged solutions of five-dimensional Chern-Simons gravity.

\end{abstract}

\pacs{Valid PACS appear here}
\maketitle

\section{Introduction}

In this note we discuss the possibility of finding a class of
black hole for which the behavior of the temperature as a function
of the horizon radius presents a {\it double peak} form.
Typically, this particular behavior leads to the presence of a
{\it plateau} in the evaporation rate, implying a drastic
evaporation for those black holes having sizes which are bounded
between the two scales where the peaks are located. We show that
these black holes actually appear as solutions of Lovelock
theory of gravity coupled to a particular non-linear
electrodynamics. The existence of such a phase behavior is due
to the fact that the two models considered here (namely Lovelock
theory and Hoffman-Infeld theory) represent short distance
corrections to both general relativity and Maxwell electrodynamics
respectively and, consequently, two peaks arise if the scale
induced by both corrections do not coincide (scale splitting).

Specifically, we study five-dimensional solutions representing
charged black holes in Hoffmann-Infeld electrodynamics within the
framework of Lovelock theory of gravity. Currently, in a
particular context, the study of the combined problem of
considering certain models of non-linear (Born-Infeld like)
electrodynamics and higher order gravitational theories acquires
importance due to the role that these theories play in low energy
string inspired models. Originally, Hoffmann-Infeld model was
proposed to avoid certain {\it pathological} features that
Born-Infeld field theory presents when spherically symmetric
static solutions are considered, as the conical singularities that
BIons present at the origin. Actually, the modification of
Born-Infeld theory presented in Ref. \cite{hi} has been shown to
lead to spherically symmetric particle-like objects whose
associated metric is regular everywhere, so avoiding the conical
singularity of the Einstein-Born-Infeld case previously studied in
Ref. \cite{h}. Nevertheless, the black hole solutions in
Einstein-Hoffmann-Infeld field theory are still singular since the
curvature diverges at the origin.

The explicit form of Hoffmann-Infeld action is presented in
Ref. \cite{hi}. This can be written as follows
\begin{equation*}
S_{HI} = -\frac {b^2}{4} \int d^5 x \sqrt{-g} \ ( 1- \eta _{(F)}- \log 
\eta _{(F)} ) \label{hi8}
\end{equation*}
with
\begin{equation}
\eta _{(F)}= \frac {b^{-2}F_{\mu \nu }F^{\mu \nu }}
{\sqrt {1+2b^{-2}F_{\mu \nu }F^{\mu \nu }}-1 } 
\label{sf}
\end{equation}
and where $b$ represents a characteristic field, analogue to that appearing in 
Born-Infeld theory. Actually, Hoffmann-Infeld model corresponds to logarithmic
modifications to a non-linear Born-Infeld-like Lagrangian, and was
originally designed in such a way that certain {\it regularity
conditions} hold for both gravitational and electric fields when
particle-like solutions are considered. In the case of the
gravitational field, the regularity condition comes from the
choice of an integration constant that amounts to state the
identity between gravitational and electromagnetic mass.

On the other hand, the short distance corrections carried by
higher order theories, as Lovelock theory of gravity,
automatically held the divergences associated with the Newtonian
term \cite{pioneros}. Hence, the finitness of the gravitational field at the
origin is guaranteed {\it ab initio}. This means that the
identification between electromagnetic and gravitational mass in
Lovelock gravity does not come from imposing the requirement for
the metric of the spacetime to be finite but by adding the
requirement that no conical singularity should exist there
\cite{afgz}.

In the following section, we describe the charged black hole solutions in five-dimensional Lovelock gravity coupled to Hoffmann-Infeld non-linear electrodynamics. In section 3, we study the thermodynamics of this solution, with particular interest focused on the evaporation phenomenon.

\section{Charged Black Hole Solutions}

The most general five-dimensional gravity action that depends on
the metric and its derivatives up to the second order and,
besides, leads to conserved field equations is the Lovelock
gravitational action. This is given by supplementing
Einstein-Hilbert action with Gauss-Bonnet terms. In lower
dimensional models ($D<5$) these terms represent topological
invariants and, hence, Lovelock gravity turns out to coincide with
general relativity. In five dimensions the action is
\begin{equation*}
S = \frac {1}{16\pi } \int d^5x \sqrt {-g}  \left( R - 2\Lambda +
\alpha (R_{\mu \nu \rho \sigma } R^{\mu \nu \rho \sigma } +
\right.
\end{equation*}
\begin{equation*}
\left. + R^2 - 4R_{\mu \nu } R^{\mu \nu} )\right) + S_{HI}
\end{equation*}
where $\alpha $ is the Gauss-Bonnet coupling constant, which
defines a length scale. Actually, this theory introduces short
distance corrections to general relativity, implying the existence
of a scale $l_{\alpha}=\sqrt{4 \alpha}$ where such corrections
turn out to be relevant.

The gravitational equations of motion resulting from $\delta S=0$
are
\begin{eqnarray*}
8\pi T_{\mu \nu } = R_{\mu \nu }-\frac{1}{2}Rg_{\mu \nu}+\Lambda g_{\mu \nu}  \\
-\alpha \left(\frac {1}{2} g_{\mu \nu} (R_{ \rho \delta \gamma \lambda}R^{\rho
 \delta \gamma \lambda }-4R_{\rho \delta }R^{\rho \delta }+R^2) \right. -   \\
\left. - 2RR_{\mu \nu}+4R_{\mu \rho}R^{\rho}_{\nu}+4R_{\rho \delta}
R^{\rho \delta}_{ \ \ \mu \nu}-2R_{\mu \rho \delta \gamma}R_{\nu}^{ \ \rho \delta \gamma}
\right)
\end{eqnarray*}
where $T_{\mu \nu}$ is the stress-tensor representing the {\it
matter-field} distribution coming from the variation $\delta
S_{HI}/\delta g^{\mu \nu}$.

Here, we are interested in static spherically symmetric solutions
satisfying the following {\it ansatz}
\begin{eqnarray*}
ds^2 &=& - g_{\alpha } (r)dt^2+ g_{\alpha }^{-1} (r)dr^2+r^2d
\theta ^2+ \\
&& + r^2 sin^2 \theta d\chi^2 +\:r^2\:sin^2 \theta\: sin^2 \chi \:d\varphi^2
\end{eqnarray*}
which, once replaced in the field equations, yields
\begin{equation}
-\frac{3}{r^3}\frac{d}{dr} \left(
r^2(1-g_{\alpha}(r))+2\alpha(1-g_{\alpha}(r))^2 \right) +2\Lambda
= 16\pi T^0_{ \ 0}\label{equa}
\end{equation}
Then, it is straightforward to prove that the following functional relation holds
\begin{equation}
g_{\alpha } (r) - g_{0 } (r) = \frac {2 \alpha}{r^2} (1-g_{\alpha
} (r))^2 \label{relation}
\end{equation}
being $g_{0 }(r)$ a spherically symmetric solution of the field
equations with $\alpha =0$, which is simply determined by solving
Einstein equations in five dimensions. The meaning of equation
(\ref{relation}) can be intuitively understood by the following
heuristic argument: let us consider the gravitational potential
$\phi_{\alpha}(r)$ defined as $g_{\alpha}(r)=1-2\phi_{\alpha}(r)$;
then, according to the relation above, the potential can be written
as $\phi_{\alpha}(r)= \phi_{0}(r)+\frac{m_{g}(r)}{r^2}$ where the
{\it mass} $m_{g}$ is due to the gravitational potential itself and
given by $m_{g} = -(l_{\alpha}\phi_{\alpha}(r))^2$.

Besides, when the total energy of the source is finite, $g_0$ can be written as
\begin{equation}
g_0 (r) = 1+\frac{16\pi}{3r^2}\int _0^r ds \ s^3 \ T^0_{ \ 0}(s) -
\frac {\Lambda}{6} r^2  \label{op}
\end{equation}
This solution amounts the choice of a null integration constant in
(\ref{equa}) (a possible term $k/r^2$ has been removed in
(\ref{op})), and implies the identification of gravitational and
electromagnetic masses. The finitness of the total energy
guarantees the Newtonian behavior at the infinity.

The non-linear electrodynamics with finite total energy are
characterized by a field scale $b$ defining the typical length
scale $l_b = (e/b)^{1/3}$, where $e$ is the charge of the object.
We will consider the electromagnetic stress-tensor for a particle-like source as having the generic form
\begin{equation}
T_0^0(r) = -\frac {b^2}{4\pi r^3} h_b(r)
\end{equation}
In the case of Hoffmann-Infeld model, the function $h_b(r)$ is given by
\begin{equation}
h_b(r) = \frac 12 r^3 \log (1+l_b^6 r^{-6}) \ ,
\end{equation}
which  corresponds to a charged particle-like source with electric field 
$E(r) 
= e /(r^{3}+l_b^6r^{-3})$. On the other hand, in the case of the Born-Infeld model the function is $h_b(r) =\sqrt {r^6+l_b^6}-r^3$. Moreover, in the generic case \footnote{which are not satisfied by the particle-like source yielding from Maxwell theory.} we will demand the following {\it finitness conditions}:
\begin{equation}
\lim _{r \to 0} \frac {1}{r^{2}} \int _0 ^r ds \ \ h_{b}(s) =
\delta<\infty  \ , \ \ \int _0 ^{\infty } ds \ \ h_{b}(s) =
\gamma<\infty \label{fin}
\end{equation}
Notice that, in the case of Hoffmann-Infeld model one finds $\gamma = \frac {\pi l_b^4}{4\sqrt{3}}$. The first condition in (\ref{fin}) is the requirement for the
metric to be finite when a particle-like solution is considered as
the source of Einstein gravity theory. On the other hand, the
second condition means that the total energy of the particle turns
out to be finite. The Born-Infeld charge fulfills both
requirements (Hoffmann studied a Born-Infeld charged black hole in
four dimensions, in the context of Einstein gravity \cite{h}).
However the Born-Infeld charge yields $\delta\neq 0$, which means
that a conical singularity remains in the metric. Hoffmann and
Infeld removed the conical singularity by modifying the
Born-Infeld electrodynamics in order to obtain $\delta=0$
\cite{hi}. However, also Lovelock gravity remove the conical
singularity, as it can be seen in the relation (\ref{relation}).
In fact, it results
\begin{equation}
g_{\alpha }(r) = 1 + \frac {r^2}{l^2_{\alpha} } +\epsilon \frac
{r^2}{l^2_{\alpha} } \sqrt { 1 +  \frac {8b^2l^2_{\alpha} }{3r^4}
\int _0 ^r ds \ h_{b}(s) + \frac {l^2_{\alpha}}{l^2_{\Lambda} }}
\label{sol}
\end{equation}
where $\epsilon = \pm 1$ and $l^2_{\Lambda } = 3/{\Lambda}$.
Thus, we find that $\displaystyle \lim _{r \to 0} g_{\alpha}(r)=1$ for any finite $\delta$.
Note that this is a charged Deser-Boulware \cite{pioneros} black hole. Solution (\ref{sol}) presents an external horizon located at $r=r_+$, with
\begin{equation}
l^2_{\alpha }+r_+^2= \frac {8b^2}{3}\int _0^{r_+} dr \ h_b (r) + \frac {r_+^4}{l^2_{\Lambda }} \label{pungas}
\end{equation}
Moreover, note that, in the large $r/l_{\alpha}$ limit, the following asymptotic behaviour is obtained
\begin{equation}
g_{\alpha }(r) = 1 +\lambda_{\epsilon} r^2 +\frac {4\epsilon e^2}{3l_b^6 r^2} \int_0^r ds \ h_b(s) + ...
\label{solita}
\end{equation}
where $\lambda _{\epsilon} = (1+\epsilon)/l^2_{\alpha}+ \epsilon/2l^2_{\Lambda}$ and the dots refer to subleading orders in $l_{\alpha}/r$ (and subleading orders in $l_{\alpha}/l_{\Lambda}$ as well). Furthermore, if the large $r/l_b$ limit of the metric is performed, we find
\begin{equation}
g_{\alpha }(r) = 1 + \lambda_{\epsilon} r^2 -\frac {2m_{\epsilon }}{\pi r^2}+ ...
\label{solas47}
\end{equation}
once the mass is accordingly identified as $m_{\epsilon }=-\epsilon
\frac{2 \pi ^2 e^2\gamma }{3l^6_b}$. Conversely, if we first take
the limit $b \to \infty$ ({\it i.e.} $l_b/r \to 0$) and then explore
the asymptotic behavior, the geometry becomes
\begin{equation}
g_{\alpha }(r) = 1 + \lambda_{\epsilon} r^2 -\epsilon \frac{e^2}{3r^4}+ ...
\label{solas}
\end{equation}
which, for instance, in the case $\epsilon =+1$ corresponds to a black hole with a dominant cosmological term $\lambda \sim l^{-2}_{\alpha}$ and a {\it wrong sign} Reissner-Nordstr\"om term $\sim -e^2/r^4$. This mimics a charged massless black hole with {\it imaginary electric charge}; though it has to be emphasized that the mechanism leading to such a metric is substantially different to the one leading to black holes with an analogous {\it tidal charge} term, cf. \cite{papadopoulos}. It is relatively simple to verify that, considering subleading effects in powers of $l_{\alpha}/l_{\Lambda}$, the mass of (\ref{sol}) is given by
\begin{equation}
m = \frac {2\pi b^2\gamma}{3}\left( 1+\sum_{n=2}^{\infty}
\frac{(2n-3)!!}{2^{n-1}(n-1)!}(-l_{\alpha}^2/l_{\Lambda}^{2})^{n-1}\right)
\end{equation}
where $l_{\alpha}<l_{\Lambda}$ and where the {\it dressing} of the
Newtonian term manifestly appears due to the presence of the
cosmological constant $\Lambda$ (see \cite{afgz}). This dressing
effect is characterized by the expansion in powers of the
dimensionless parameter $l^2_{\alpha}/l^2_{\Lambda}$. Moreover,
the specific value of the first term in such expansion is the one
required for the metric to be regular at $r=0$.

On the other hand, for the specific value $l^2_{\alpha } =
-l^2_{\Lambda }$ ($\alpha >0$, $\Lambda <0$), the solution takes a
rather different form;

namely
\begin{equation}
g_{\alpha }(r) = 1 + \frac {r^2}{l^2_{\alpha} } + c(r)
\label{sol3}
\end{equation}
with the function
\begin{equation}
c^2(r)= \frac {8b^2}{3l^2_{\alpha }} {\int _0 ^{r} ds \ \ h_b (s)
}  \label{tenuza}
\end{equation}

\begin{figure}
\begin{center}
\includegraphics[width=9cm,angle=0]{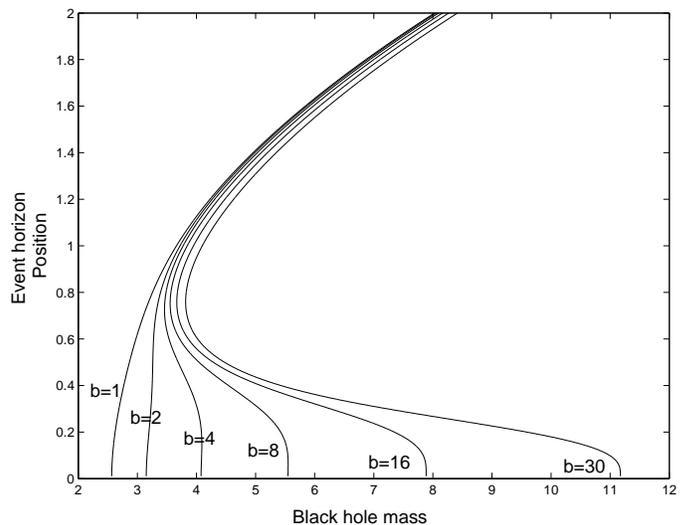}
\caption[]{The position of the event horizon as a function of the black hole
mass for different values of $b$ [ $\alpha=0.5$, $e=1$, $\Lambda=0$ ].}
\label{fig9}
\end{center}
\end{figure}

\begin{figure}
\begin{center}
\includegraphics[width=9cm,angle=0]{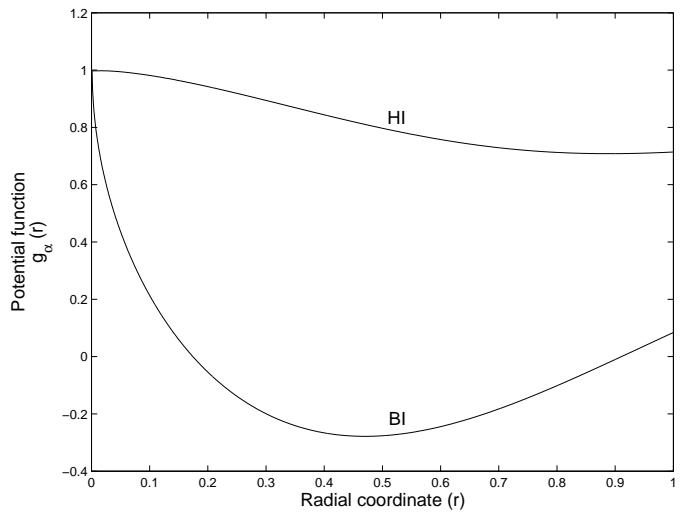}
\caption[]{The (potential) function $g_{\alpha}(r)$ for both
Hoffmann-Infeld (HI) and Born-Infeld (BI) solutions [
$\alpha=0.1$, $e=1$, $\Lambda=0$].} \label{origen}
\end{center}
\end{figure}

This metric corresponds to an asymptotically Anti-de Sitter space
in five dimensions. This is due to the finite value $c^2(\infty
)=\frac {8b^2\gamma}{3l^2_{\alpha }}$ at infinity. Geometries
(\ref{tenuza}) present event horizons and are closely related to
the black hole solutions of Chern-Simons gravitational theory.

In particular, for the Lovelock-Hoffmann-Infeld black hole, the function $h_b(r)$ clearly satisfies the finitness conditions described above. In this case, the charged black hole solution (\ref{sol}) shares
several properties with the one built for the Born-Infeld model,
{\it e.g.} the existence of charged black holes with a unique
horizon. However, the horizon structure of both theories is
certainly different (cf. Fig.\ref{fig9} and the analogue presented
in Ref. \cite{afg}), {\it e.g.} the fact that the internal radius
$r_-$ decreases when the mass $m$ increases (for a fixed charge
$e$) is strongly more evident for the case of Hoffmann-Infeld
black hole. Moreover, by comparing the solution (\ref{sol}) and
the one corresponding to the Lovelock-Born-Infeld black holes
\cite{afg,afg2,afg3,afg4,papadopoulos,afgz}, it is feasible to verify that in both
theories $g_{\alpha}(r) $ can be set to $1$ at the origin $r=0$.
Nevertheless, in the case of Lovelock-Hoffmann-Infeld solutions,
we find that $\frac{dg_{\alpha}(r)}{dr}$ vanishes in the limit
$r\rightarrow 0$ whereas it goes to $-\infty$ for
Lovelock-Born-Infeld black holes (see Fig. \ref{origen} which
manifests the difference between both solutions at the origin).
This is because the short distance corrections to Maxwell theory
involved in Hoffmann-Infeld are, in some sense, stronger than the
ones corresponding to Born-Infeld model.

\section{Thermodynamics}

Now, the question arises as to what are the thermodynamical
properties of this charged black hole solution. Certainly, it is
known that black holes in Lovelock gravity present special
features which are not shared with their analogues in Einstein
gravity theory; namely, these black holes typically have an
infinite lifetime, present a positive specific heat for small
radii, their isotermal graphs for the charged cases turn out to be
rather different  \cite{afg}, violate the Bekenstein's area
formula \cite{afg2,afg3} and the temperature formula in the
general case presents an additional term which identically
vanishes for $D=5$. Here, we show that, besides these remarkable
aspects, the Lovelock-Hoffmann-Infeld black holes present a {\it
plateau} in the evaporation rate as a function of the horizon
radius $r_+$. This aspect implies that the solutions
charged under the Hoffmann-Infeld electrodynamics evaporate
drastically when they have middle sizes within the range of scales
where the specific heat is negative. Eventually, these black holes
end up in a stable phase (region of positive specific heat) and their lifetimes result infinite. This
can be intuitively inferred from the fact that the caloric curve
presents a double peak form for certain tunning of the parameters.

\begin{figure}
\begin{center}
\includegraphics[width=9cm,angle=0]{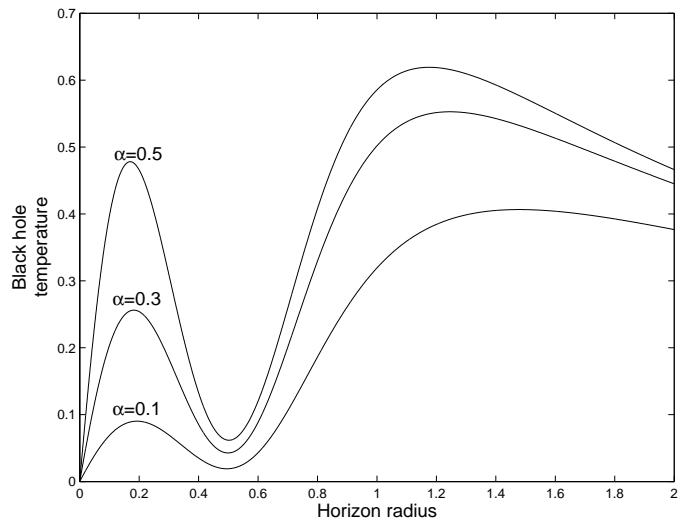}
\caption[]{Black hole temperature as a function of the horizon
radius [ $b=2$, $e=1$ ]; representing the caloric curve.}
\label{fig1a}
\end{center}
\end{figure}

\begin{figure}
\begin{center}
\includegraphics[width=9cm,angle=0]{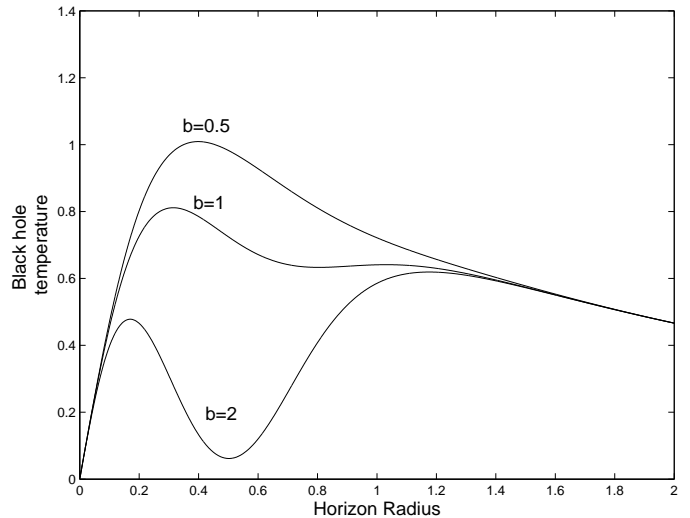}
\caption[]{Black hole temperature as a function of the horizon
radius [ $\alpha=0.05$, $e=1$ ]; representing the caloric curve.}
\label{temp3}
\end{center}
\end{figure}

Let us begin by writting down the temperature for the case of
vanishing cosmological constant; namely
\begin{equation}
T=\frac {1}{4\pi }\frac {dg_{\alpha }(r)}{dr} \Big{|}_{r=r_+}=\frac 
{r_+-\frac{2b^2}{3}h_b(r_+)}{2\pi (l_{\alpha }^2+r_+^2)} \label{T}
\end{equation}
whose typical form is described in Fig. \ref{fig1a} and
\ref{temp3}. The expression above corresponds to the temperature of solution (\ref{sol}) with $\epsilon =-1$ and $\Lambda =0$; this solution is asymptotically flat, as it can be verified by means of the expansion (\ref{solas47}). Actually, the regime of general relativity is recovered in the limit $\l_{\alpha }/r_+ \to 0$, where the expression for the temperature takes the form $T \sim 1/r_+$.

We also notice that the specific heat changes its sign due to the short
distance corrections imposed by both Lovelock and Hoffmann-Infeld
models. The sign of the specific heat enables one to infer which
are the regions of thermodynamical stability (where the black hole can be in thermal equilibrium with the enviroment). Consequently, the
evaporation rate of these charged solutions is obtained from
(\ref{T}) by integrating over the energy flux. This is done by making use of the 
Stefan-Boltzmann law in five dimensions; namely $\frac {dM_s}{d\tau} \sim T^5$, 
being $M_s$ the surface energy density ($M_s \sim m / r_+^3$). Then, by using that 
the 
(three-dimensional) surface of the horizon is given by $\frac {\pi }{2} 
r_+^3$ and the expression of the temperature is given by (\ref{T}) we can integrate over the black hole size in order to obtain the evaporation time
\begin{equation}
\tau \sim \int ^{r_0}_{r_+} ds \frac {(l^2_{\alpha }+s^2
)^5}{s^3(s-\frac 23 b^2h_b(s))^4}
\end{equation}
where (\ref{pungas}) was also taken into account. This corresponds to the time required for a black hole to evaporate, starting with the initial size $r_0$ and ending with a size $r_+$. Notice that $r_+$ is a monotonic function of the mass (energy) $m$. The
symbol $\sim$ stands in the formula above because of the presence of a 
positive
multiplicative constant which is given in terms of the (inverse
of) Stefan-Boltzmann constant in five dimensions.

\begin{figure}
\begin{center}
\includegraphics[width=9cm,angle=0]{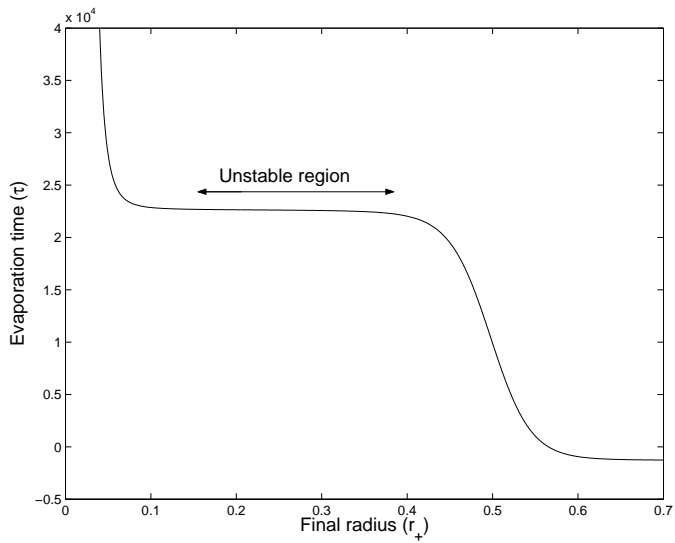}
\caption[]{Evaporation time $\tau$ as a function of the final
radius $r_+$ [$\alpha=0.05$, $e=1$, $b=2$]. A {\it plateau}
manifestly appears in the evaporation rate. The graph has been
normalized by means of an appropriate redefinition of the
Stefan-Boltzmann constant.} \label{tiempo}
\end{center}
\end{figure}

This result leads to observe the presence of the {\it plateau}
which is studied in Fig. \ref{tiempo}, showing a rapid
transition between the two scales where the first maximum and the
local minumum of Fig. \ref{fig1a} are located \footnote{Let us
remark the qualitative analogy existing between the caloric curve
of the Lovelock-Hoffmann-Infeld black holes and those
corresponding to the models with long-range interactions in
condensed matter. In fact, it is well known that certain
quasi-stationary states of those models exhibit a similar (double)
change of sign in the specific heat and actually are qualitatively
similar to the profile displayed in Fig. \ref{temp3} (for
instance, cf. \cite{condmat} and references therein). The caloric
curve obtained here is reminiscent of that.}. The evaporation rate is usually displayed by analyzing the quantity ${dm}/{d\tau }$; let us notice that the Figure \ref{temp3} (showing the time $\tau $ required to reach a size $r_+$) basically gives the same information: This is because the {\it plateau} of the graph precisely corresponds to those scales for which the transition (evaporation) is abrupt and, hence, the quantity ${dm}/{d\tau }$ would present a peak precisely located in that region. Moreover, since we are interested in studying the scales where such an abrupt transition occurs, we find Figure \ref{temp3} convenient because it manifestly shows those scales within which such a drastic effect takes place.

An interesting analysis of the black hole thermodynamics in
Einstein-Gauss-Bonnet gravity and Chern-Simons gravity was
recently performed in references \cite{zanelli,a,aa,b,c,zanelliz,olea}.
To make contact with the Chern-Simons gravity, let us consider
again the case $l_{\alpha}^2 = - l^2_{\Lambda}$, for which the
formula of the temperature as a function of the horizon radius
$r_+$ acquires a dominant linear term; namely
\begin{equation}
T=\frac {r_+}{2\pi l^2_{\alpha}}-\frac {b^2h_b(r_+)}{3\pi
(l^2_{\alpha }+r_+^2)}
\end{equation}
This diverges in the limit $r_+/l_{\alpha } \to 0$.

Summarizing, Lovelock theory of gravity in higher dimensions
introduces short distance corrections to general relativity due to
Gauss-Bonnet terms which, in addition to Einstein-Hilbert action,
have to be taken into account in the most general theory of
gravity. These terms, corresponding to Lanczos quadratic gravity
for $D=5$, are such that the mentioned short distance effects
imply substantial differences with respect to the black hole
thermodynamics of general relativity; these were listed above. In
addition, we discussed here how the charged black holes in
Lovelock five dimensional gravity coupled to non-linear
Hoffmann-Infeld electrodynamics present other interesting features
like the existence of the double peak profile in the caloric
curve, leading to a particular evaporation effect for which two
different thermodynamically stable regions do exist. The black holes evaporate
drastically for certain sizes that are bounded by two critical
radii; this region corresponds to the one where the specific heat
is negative. Eventually, a final phase is reached and the black
holes become eternal; the explicit computation of the evaporation
rate leads to an infinite lifetime as result. Because of the
particular profile of the caloric curve, the evaporation
phenomenon described here is qualitatively different to the one
corresponding to the five-dimensional Lovelock-Born-Infeld
solutions, cf. \cite{afg,afg2,afg3,afg4,papadopoulos,afgz}.

Furthermore, the analysis of the static spherically symmetric
solution presented in this note is general enough as to be
suitable for adaptation to the case of Lovelock black holes
charged under a quite generic model of non-linear electrodynamics.
In particular, it would be relatively easy to extend it to those
models of electrodynamics leading to regular black holes
\footnote{For such models, the static black hole solutions present
a change of the topology of the causal structure and the finite
curvature at the origin is allowed because of the non-existence of
a non-compact Cauchy surface.} in Einstein gravity
\cite{eloy,eloy2,eloy3,eloy4,eloyz}. The analysis of these geometries within
the framework of Lovelock gravity could be an interesting subject
for further study.

Before concluding, let us make brief remark on the higher dimensional case. Certainly, the five-dimensional case presents a special feature: the fact that expression for the temperature (\ref{T}) acquires an additional term in $D$ dimensions, which is proportional to $(D-5)$. This is precisely why previous papers of the subject (see for instance \cite{afg}) considered the case $D=5$ as an special one. However, it is also true that, besides of that, several qualitative aspects of the termodynamics of the $5D$ Lovelock black holes are shared with their higher dimensional analogues: For instance, this is the case of the change of the sign of the specific heat at short distances and the existence of infinite lifetime remnants. Then, similar features to those analyzed in this paper are expected to be valid in the $D$-dimensional Lovelock-Hoffmann-Infeld black holes.

\acknowledgments

R.F. was supported by Universidad de Buenos Aires (UBACYT X103) and
Consejo Nacional de Investigaciones Cient\'{\i}ficas y T\'ecnicas
(Argentina). G.G. was supported by Universidad de Buenos Aires. G.G.
specially thanks the group of Centro de Estudios Cient\'{\i}ficos
CECS; in particular, Eloy Ay\'on Beato, Rodrigo Olea and Jorge Zanelli for
fruitful conversations on related topics. The authors also thank E.
Calzetta and M. Ison for pointing out interesting references.

\end{document}